# Neutrino geophysics with KamLAND and future prospects

S. Enomoto [a,*], E. Ohtani [b], K. Inoue [a], A. Suzuki [a]

[a] *Research Center for Neutrino Science, Department of Physics, Graduate School of Science, Tohoku University, Sendai 980-8578, Japan*

[b] *Institute of Mineralogy, Petrology and Economic Geology, Graduate School of Science, Tohoku University, Sendai 980-8578, Japan*

**Abstract**

The Kamioka liquid scintillator anti-neutrino detector (KamLAND) is a low-energy and low-background neutrino detector which could be a useful probe for determining the U and Th abundances of the Earth. We constructed a model of the Earth in order to evaluate the rate of geologically produced anti-neutrinos (geoneutrinos) detectable by KamLAND. We found that KamLAND can be used to determine the absolute abundances of U and Th in the Earth with an accuracy sufficient for placing important constraints on Earth's accretion and succeeding thermal history. Within the uncertainty of the measurement, the present observation of geoneutrinos with KamLAND is consistent with our model prediction based on the bulk silicate Earth (BSE) composition. If a neutrino detector were to be built in Hawaii, where effects of the continental crust would be negligible, it could be used to estimate the U and Th content in the lower mantle and the core. Our calculation of the geoneutrino event rate on the Earth's surface indicates that geoneutrino observation can provide key information for testing the current models of the U and Th content and distribution in the Earth.

# Introduction

It is widely accepted that radiogenic heat contributes a large part to the Earth's heat budget. Due to the direct relation between the number of radioactive decays in the Earth and the neutrino flux from the decays, neutrinos are expected to provide fruitful information on Earth's energetics. The Earth's dynamical activities, such as plate tectonics, are powered by the Earth's heat generation and transportation processes. The heat flux coming from the Earth's interior is about 87 mW/m$^2$ or $44.2 \pm 1.0$ TW (1 TW $=10^{12}$ W) in total [1]. Despite the small quoted error, a more recent re-evaluation of the same data has led to a lower figure of $31 \pm 1$ TW [2]. The total heat flux is an important constraint for thermal models of the Earth. Based on our knowledge of the absolute abundances of radioactive isotopes such as $^{238}$U and $^{232}$Th in the Earth, we can evaluate the Earth energy budget and determine whether the Earth is heating up, in thermal equilibrium, or cooling down.

Based on chondritic abundances of U and Th, and cosmochemical consideration of the volatility of K, the current model of the bulk silicate Earth (BSE) [3] gives heat generation of 8 TW by $^{238}$U and $^{235}$U, 8 TW by $^{232}$Th, and 3 TW by $^{40}$K, resulting in total of 19 TW for the radiogenic heat generated in the Earth's interior. The discrepancy between the total surface heat flux (44 TW or 31 TW) and the total radiogenic heat (19 TW) is a matter of debate. It is generally assumed that the heat flux from the Earth's core is 5-10 TW, which is related to the thermal history and evolution of the core.

Use of neutrinos to study Earth science was first suggested by Marx [4,5], Markov [6] and Eder [7] in 1960's, and then reviewed several times by several authors [8-13]. The Kamioka liquid scintillator anti-neutrino detector (KamLAND) is the first detector sensitive enough to detect geologically produced anti-neutrinos (geoneutrinos). Despite the extremely small cross section for neutrino interactions, the KamLAND collaboration recently reported the first experimental results [14].

The purpose of this study is to provide a certain framework for possible application of neutrinos in geophysical research, and also to foresee future sensitivities of experimental studies including KamLAND.

**Radiogenic heat generation and geoneutrino flux**

$^{238}$U, $^{235}$U and $^{232}$Th generate radiogenic heat via a series of alpha and beta decays. $^{40}$K generates radiogenic heat via either beta decay or electron capture with branching ratios of 0.893 and 0.107, respectively. In addition to a daughter nucleus, each beta decay produces an electron and an anti-neutrino. Electron capture produces a neutrino and a daughter nucleus. These decays can be summarized by

$$\begin{aligned}
^{238}\text{U} &\rightarrow\, ^{206}\text{Pb} + 8\,^{4}\text{He} + 6e^{-} + 6\bar{\nu}_e + 51.7\,\text{MeV} \\
^{235}\text{U} &\rightarrow\, ^{207}\text{Pb} + 7\,^{4}\text{He} + 4e^{-} + 4\bar{\nu}_e + 46.0\,\text{MeV} \\
^{232}\text{Th} &\rightarrow\, ^{208}\text{Pb} + 6\,^{4}\text{He} + 4e^{-} + 4\bar{\nu}_e + 42.7\,\text{MeV} \\
^{40}\text{K} &\rightarrow\, ^{40}\text{Ca} + e^{-} + \bar{\nu}_e + 1.31\,\text{MeV} \\
^{40}\text{K} + e^{-} &\rightarrow\, ^{40}\text{Ar} + \nu_e + 1.51\,\text{MeV}
\end{aligned} \qquad (1)$$

Anti-neutrino luminosity $L_{\bar{\nu}}$ (number of anti-neutrino emissions per unit time) and heat generation $Q_{\text{heat}}$ of $^{238}$U, $^{235}$U, $^{230}$Th and $^{40}$K can be directly calculated from their mass $M$ by

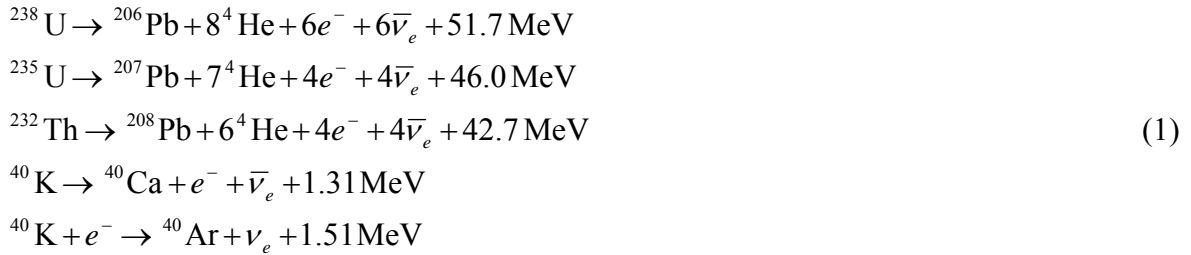

$$\begin{aligned}
^{238}\text{U}:\ & L_{\bar{\nu}}[1/\text{sec}] = 7.46 \times 10^{7} \cdot M[\text{kg}] = 7.84 \times 10^{11} \cdot Q_{\text{heat}}[\text{W}] \\
^{235}\text{U}:\ & L_{\bar{\nu}}[1/\text{sec}] = 3.20 \times 10^{8} \cdot M[\text{kg}] = 5.67 \times 10^{11} \cdot Q_{\text{heat}}[\text{W}] \\
^{232}\text{Th}:\ & L_{\bar{\nu}}[1/\text{sec}] = 1.62 \times 10^{7} \cdot M[\text{kg}] = 6.18 \times 10^{11} \cdot Q_{\text{heat}}[\text{W}] \\
^{40}\text{K}:\ & L_{\bar{\nu}}[1/\text{sec}] = 2.31 \times 10^{8} \cdot M[\text{kg}] = 8.18 \times 10^{12} \cdot Q_{\text{heat}}[\text{W}]
\end{aligned} \qquad (2)$$

where the neutrino kinetic energy is subtracted from $Q_{\text{heat}}$, since almost all this energy escapes the Earth. In order to calculate the energies taken by neutrinos on beta decays, the allowed beta transition formula with the Fermi correction is used except for $^{40}$K and $^{210}$Bi in the $^{238}$U series; the 3rd unique forbidden transition formula is used for $^{40}$K, and the spectrum tabulated in [15] is used for $^{210}$Bi. The isotope decay data used in this calculation is taken from [16]. Using natural abundance of these isotopes, the relation (2) can be rewritten for natural elements;

$$^{\text{nat}}\text{U}: \quad L_{\bar{\nu}}[1/\text{sec}] = 7.64 \times 10^7 \cdot M[\text{kg}] = 7.75 \times 10^{11} \cdot Q_{\text{heat}}[\text{W}]$$
$$^{\text{nat}}\text{Th}: \quad L_{\bar{\nu}}[1/\text{sec}] = 1.62 \times 10^7 \cdot M[\text{kg}] = 6.18 \times 10^{11} \cdot Q_{\text{heat}}[\text{W}] \quad (3)$$
$$^{\text{nat}}\text{K}: \quad L_{\nu}[1/\text{sec}] = 2.70 \times 10^4 \cdot M[\text{kg}] = 7.98 \times 10^{12} \cdot Q_{\text{heat}}[\text{W}]$$

Since the natural abundance of $^{235}$U is 0.72%, $^{235}$U contributes only 3.0% in neutrino luminosity and 4.1% in heat production of $^{\text{nat}}$U.

For each isotope the neutrino flux at a position $\vec{r}$ can be calculated from the isotope distribution (isotope mass per unit rock mass) $a(\vec{r}')$ by integrating the contribution over the entire Earth,

$$\frac{d\Phi(E_{\nu},\vec{r})}{dE_{\nu}} = A \frac{dn(E_{\nu})}{dE_{\nu}} \int_{\oplus} d^3\vec{r}' \frac{a(\vec{r}')\rho(\vec{r}')P(E_{\nu},|\vec{r}-\vec{r}'|)}{4\pi|\vec{r}-\vec{r}'|^2}, \quad (4)$$

where $A$ is the decay rate per unit mass, $dn(E_{\nu})/dE_{\nu}$ is the energy spectrum of neutrinos per decay, $\rho(\vec{r}')$ is the rock density, and $P(E_{\nu},|\vec{r}-\vec{r}'|)$ is the neutrino survival probability after traveling from the source position $\vec{r}'$ to the detector position $\vec{r}$. The neutrino survival probability is given by a well-established formula,

$$P(E_{\nu},L) \cong 1 - \sin^2 2\theta_{12} \sin^2\left(\frac{1.27 \Delta m_{12}^2[\text{eV}^2]L[\text{m}]}{E_{\nu}[\text{MeV}]}\right) \quad (5)$$

where $L = |\vec{r}-\vec{r}'|$. The neutrino oscillation parameters $\sin^2 2\theta_{12} = 0.82^{+0.07}_{-0.07}$ and $\Delta m_{12}^2 = 7.9^{+0.6}_{-0.5} \times 10^{-5} \text{ eV}^2$ used in the model calculation are determined by the KamLAND experiment combined with solar neutrino experiments [17]. Corrections from the "matter effect" are found to affect at about 1% and are ignored in (5).

**Geoneutrino detection with KamLAND**

KamLAND detects electron anti-neutrinos via neutron inverse beta-decay,

$$\bar{\nu}_e + p \rightarrow e^+ + n \tag{6}$$

which has a 1.8 MeV neutrino energy threshold, and a well established cross-section [18]. The maximum antineutrino energy from $^{40}$K and $^{235}$U decays are 1.3 MeV and 1.4 MeV, respectively, both of which are below the inverse beta-decay threshold. The number of geoneutrino events from the reaction for the $^{238}$U and $^{232}$Th antineutrinos are calculated by

$$N = N_{proton} \cdot \tau \cdot \varepsilon \cdot \int dE_\nu \sigma(E_\nu) \frac{d\Phi(E_\nu)}{dE_\nu} \tag{7}$$

where $N_{proton}$ is the number of target protons, $\tau$ is the detector exposure time, $\varepsilon$ is the detection efficiency, and $\sigma(E_\nu)$ is the cross section of the reaction (6). With 1-year exposure of $10^{32}$ target protons assuming 100% efficiency, 1 $^{238}$U-neutrino event corresponds to a flux of $7.67 \times 10^4 \, \text{cm}^{-2} \, \text{sec}^{-1}$, and 1 $^{232}$Th-neutrino event corresponds to a flux of $2.48 \times 10^5 \, \text{cm}^{-2} \, \text{sec}^{-1}$.

## A reference Earth model and geoneutrino flux at the KamLAND site

In order to provide a starting point, we construct a reference Earth model based on the bulk silicate Earth (BSE) composition given by McDonough [3] and a crustal composition model given by Rudnick & Fountain [19]. Following common practice, we assume that the core does not contain U and Th. The mantle composition is assumed to be uniform and is obtained by subtracting the crustal composition from the BSE composition. The effects of all these assumptions are discussed later. The reference Earth model contains crustal thickness and sediment thickness taken from the CRUST 2.0 data set [20] and Laske *et al.* [21], respectively, and a density profile given by PREM [22]. The geoneutrino flux at the KamLAND site ($36°25'36"\text{N}, 137°18'43"\text{E}$, 358m elevation from the sea level) is calculated by performing numerical integration of equation (4). Table 1 summarizes the reference Earth model and Table 2 shows the calculated neutrino flux at the KamLAND site. The total $^{238}$U and $^{232}$Th geoneutrino fluxes are calculated to be $2.34 \times 10^6$ cm$^{-2}$ sec$^{-1}$ and $1.99 \times 10^6$ cm$^{-2}$ sec$^{-1}$, respectively, corresponding to event rates of 30.5 and 8.0 per $10^{32}$ target protons per year.

**Global geophysics: flux response coefficient and flux response chart**

In order to separate geochemical uncertainties from the flux estimation, we define geochemical-model-independent response coefficients, $R_X$, as a ratio of flux and its source mass;

$$R_X = F_X / M_X \qquad (8)$$

where $F_X$ is the geoneutrino flux at a observation point from source in reservoir X, and $M_X$ is the mass of source in the reservoir X. Naturally $R_X$ is independent of the source concentration in the reservoir. Although response coefficients are primarily defined for each elemental reservoir in the reference Earth model, their combination is also useful and can be directly obtained. For example, the whole mantle response coefficient is a combination of the upper mantle (UM) and the lower mantle (LM) response coefficients, obtained by $R_{\text{Mantle}} = (F_{\text{UM}} + F_{\text{LM}})/(M_{\text{UM}} + M_{\text{LM}})$. In this case, $R_{\text{Mantle}}$ is independent from the total mass in the mantle ($M_{\text{UM}} + M_{\text{LM}}$) but is dependent on the ratio between the upper and lower mantle ($M_{\text{UM}}/M_{\text{LM}}$), representing reservoirs' internal structure. Table 2 shows the response coefficients of each elemental reservoir for observation with KamLAND.

The relation between geochemical models of reservoirs and the geoneutrino flux at a detector position can be visualized by plotting the (mass, flux) pairs of reservoirs on a 2-D plane, namely flux response chart. On this chart, response coefficients are expressed as slopes of a proportional line (namely response line) of $F_X = R_X \times M_X$, and alternating geochemical models (i.e., source mass in reservoirs) moves the corresponding (mass, flux) points along their response lines. Combination of reservoirs is expressed by a simple vector sum of corresponding points, and geochemical constraints such as the total amount of U and Th (e.g. the BSE constraint) are intuitively incorporated on the chart as demonstrated below.

Figure 1 shows the response chart for three representative crustal composition models, illustrating the dependence of the total flux on the crustal model. In these models, the total U and Th mass is constrained by the BSE ratios, and the mantle composition is inferred from the BSE composition and the crustal composition, under the assumption of the uniform mantle model. The chart illustrates the connection among these models, assumptions and uncertainties with relation to the total flux.

Figure 2 shows the response chart illustrating the dependence of the total flux on the mantle models, given a crustal model and the BSE constraint. The response lines of the upper and lower mantle illustrate how re-distribution of U and Th between the upper and lower mantle affects the total flux.

Since modeling of the upper/lower crust, upper/lower mantle, bulk Earth etc. are strongly connected to each other, and geochemical data rarely comes with error estimations, illustration by response charts helps understand outline of the relations and provides a way to incorporate the geoneutrino observation into Earth modeling.

## Effects of local geology

In our reference model, about 25% of the total geoneutrino flux originates within 50 km of KamLAND, and about 50% comes within 500 km radius. Since the near-field region makes a large contribution to the geoneutrino flux, local variations of U and Th concentrations may considerably influence the total geoneutrino flux, while not significantly affecting the global geo-physics. Here we evaluate the effects of possible local heterogeneities near KamLAND.

### *1. Crust in Japanese Island Arc and beneath Sea of Japan*

The continental crust surrounding KamLAND is Island Arc and Forearc crust, whose compositions might be different from the average continental crust composition. Based on our reference model and the crustal type map given by the CRUST 2.0 data set [20], it is found that the Island Arc crust and the Forearc crust contribute 60% and 15% of the total flux from the continental crust, respectively. A survey of the Japanese Island Arc geochemistry [23] shows a depletion in incompatible elements including U and Th compared to the average continental crust, reporting the average U and Th concentrations in the Japanese Islands to be 2.32 ppm and 8.3 ppm, respectively. Substituting the Island Arc and Forearc composition in the reference model with these values reduces the total U and Th geoneutrino flux by 6.4% and 8.4%, respectively.

Although the crust beneath the Sea of Japan is usually classified as oceanic, its composition is likely to be different from the typical oceanic crustal composition due to the fact that fragments of the continental crust are found in the crust beneath the Sea of Japan. In our reference model, the crust beneath the Sea of Japan contributes less than 0.1% to the total flux. By varying the composition of the crust below the Sea of Japan from that of typical oceanic crust to that of typical continental crust, the uncertainty in the total geoneutrino flux due to the uncertain crustal type is estimated to be less than 2%.

In our reference model, the composition of oceanic sediment is taken from the analysis of subducting sediment [24]. The thick sediment on the floor of the Sea of Japan, East China Sea and Sea of Okhotsk contains a large fraction of continent-derived sediment. By varying the composition of the sediment on the floor of the Sea of Japan from that of typical oceanic sediment to that of typical continental sediment, the uncertainty in the total geoneutrino flux due to the uncertain sediment composition on the floor of the Sea of Japan is estimated to be less than 0.36%.

### 2. *Subducting Plate and Accumulated Slab beneath Japan*

The geoneutrino flux from the subducting plate is calculated based on the crustal model given by Zhao *et al.* [25]. The thickness of the oceanic crust in the plate may be conservatively set to 10 km. With the normal mantle composition, the periodite lithospheric mantle component in subducting plate contributes 0.025% of the total geoneutrino flux. If we alter the composition to be the same as the oceanic crust, the total geoneutrino flux increases by 0.21% and 0.11% for U and Th, respectively.

Mantle tomography [26] shows a high seismic wave velocity anomaly beneath Japan, which suggests the accumulation of a cold slab. The region with a velocity anomaly greater than +1% contributes 10.9% of the geoneutrino flux from the whole mantle. By assuming that the slab, or the bulk subducting plate, consists of 10% of the oceanic crust (MORB + sediment) and 90% of the mantle, we estimate the total geoneutrino flux increases by 2.1% and 1.0% for U and Th, respectively.

### *3. Geology of the Japanese Islands and near the Kamioka mine*

Variation in the distribution of U and Th in Japanese Islands can be seen by integration of information on distribution of rock types from the geological map of Japan with data from a large-scale geochemical study of Japanese Islands. Geological Survey of Japan (GSJ) publishes a geological map of Japan with 165 geological groups [27]. Togashi *et al.* [23] merged the 165 geological groups defined in the GSJ geological map into 37 geological groups, and collected 166 rock samples in order to obtain representative coverage of the rock varieties and abundances in each group. Using this approach the surface exposure weighted average concentration is found to be 2.32 ppm and 8.3 ppm for U and Th, respectively.

For an estimation of how the surface geology affects the total geoneutrino flux at the KamLAND site, we assume that the surface exposed geology extends to 5 km in depth, which is generally verified from refraction seismology. With the surface exposure weighted average concentration, geoneutrino flux from this area is calculated to be 16% of the total flux. The effect of surface geology heterogeneity is studied by assigning representative rock composition to each geological group, as shown in Figure 3. If we take the average of all rock samples in a geological group as its representative composition, the geoneutrino flux from this region is changed by –20.3% and +1.0% for U and Th, respectively. If we replace the representative composition of some surrounding regions to the average of rock samples only taken around the KamLAND site, the flux for U and Th is changed by –3.8% and –14.7%, respectively. Looking at the flux variation among these different U and Th distribution models, we conclude that the flux from this region has an error of approximately 20%. This corresponds to an uncertainty of 3.2% in the total geoneutrino flux.

Although contribution of the Kamioka Mine area to the total flux is estimated to be about 1%, there is a possibility that a small-scale enrichment in U or Th, such as ore deposits, may significantly affect the total geoneutrino flux. Based on geological studies of the Kamioka mine, several rock samples were collected and analyzed [28]. The result shows an overall trend of U and Th depletion compared with the averages of the Japanese Islands, and no unusual excess in the concentration of these elements is observed.

Explorations of Uranium mines over the Japanese Islands was conducted by Geological Survey of Japan (GSJ). Taking into account that the total known Uranium deposit in Japan [29] is 8 kton, a 100 kton scale Uranium deposit, which is the size of the world largest known deposit, may be a reasonable upper limit for a possible Uranium deposit beneath the Kamioka area. The neutrino flux from a 100 kton Uranium deposit located by 1 km away from a detector is $6 \times 10^{4}$ cm$^{-2}$sec$^{-1}$, which is less than 3% of the total flux expected at the KamLAND site. Therefore it is unlikely that undiscovered huge Uranium deposits make significant contribution.

**Other sources of uncertainties**

At present, besides geological effects, the largest uncertainty source is the neutrino oscillation parameters used in equation (5). Due to vast distribution of the neutrino sources, the oscillatory behavior of $P(E_\nu, L)$ in equation (4) is averaged out by the integration. The approximation $P(E_\nu, L) \cong 1 - 0.5\sin^2 2\theta$ is found to affect accuracy of the integration result with a numerical analysis at less than 1%. The current best estimation of $\sin^2 2\theta$ is $0.82 \pm 0.07$ [17], thus the oscillation parameter uncertainty makes 6% uncertainty in flux estimation.

The reference Earth model is based on crustal thickness model given in the CRUST 2.0 dataset [20], which provides a crustal thickness map covering all Earth surface with $2° \times 2°$ resolution. This resolution corresponds to a tile of 180 km East-West and 220 km North-South around the Japan Islands. A detailed crustal thickness map of the land area of the Japanese Islands was constructed by Zhao *et al*. [25]. Although the map does not include the sea area which also makes significant contribution, it is tentatively used to estimate the effect of finite resolution of the crustal model. By partially replacing the CRUST 2.0 crustal thickness data with the fine resolution data, the total geoneutrino flux is reduced by 4%.

**The global geophysics, local geology and uncertainties**

As demonstrated above, uncertainties due to local geology, neutrino oscillation parameters and finite resolution of the crustal thickness data are less than 10%, which needs to be compared with the uncertainty in global geophysical and geochemical models of crust and mantle, such as the BSE model uncertainties (20% in U and 15% in Th), crustal models and mantle models. Constraining these non-global uncertainties allows the KamLAND observations to be used for evaluating the global geophysical properties, such as the global heat budget.

## KamLAND Observation

The KamLAND group has recently released the first results from a search for geoneutrinos with $7.09 \times 10^{31}$ target proton-years exposure [14]. Their spectrum analysis resulted in $28^{+16}_{-15}$ geoneutrino events, which corresponds to flux of $6.4^{+3.6}_{-3.4} \times 10^6 \, \text{cm}^{-2}\text{sec}^{-1}$, and the 99% CL upper limit is $16.2 \times 10^6 \, \text{cm}^{-2}\text{sec}^{-1}$. KamLAND does not have good sensitivity to the $^{232}$Th/$^{238}$U ratio, and a cosmochemically determined Th/U ratio, 3.9, is used in the analysis.

One of the primary interests of geoneutrino observation is to determine the Earth's global heat budget (i.e., test of the BSE model). The dependence of the flux on the crustal and mantle modeling (including some extreme cases) is about the same order of magnitude as the current uncertainty of the BSE composition. Based on our reference Earth model, the geoneutrino flux observed with KamLAND $F_{\text{U+Th}}$ is related to the total $^{238}$U+$^{232}$Th mass in the Earth $M_{\text{U+Th}}$ and heat generation $Q_{\text{U+Th}}$ with response coefficients listed in Table 2 as

$$Q_{\text{U+Th}}[\text{TW}] = \frac{M_{\text{U+Th}}[\text{kg}]}{2.49 \times 10^{16}} = \frac{F_{\text{U+Th}}[\text{cm}^{-2}\text{sec}^{-1}]}{2.70 \times 10^5} \tag{9}$$

Therefore the observed geoneutrino flux corresponds to $23^{+13}_{-12}$ TW of the total radiogenic heat by $^{238}$U and $^{232}$Th decays, assuming that we scale concentrations of U and Th in all reservoirs equally. If we fix the crustal composition and parameterize the mantle composition as illustrated in Figure 1 (thin solid black lines), the relation becomes

$$Q_{\text{U+Th}}[\text{TW}] = \frac{M_{\text{U+Th}}[\text{kg}]}{2.49 \times 10^{16}} = 6.58 + \frac{(F_{\text{U+Th}}[\text{cm}^{-2}\text{sec}^{-1}] - 31.7 \times 10^5)}{1.22 \times 10^5} \tag{10}$$

In this case the observed geoneutrino flux corresponds to $33^{+30}_{-28}$ TW of the total radiogenic heat. One can study the relations between the observed flux and heat generation under any other Earth models using the response coefficients and response charts in a similar way.

The 99% CL flux upper limit is 3.8 times higher than the predicted flux based on the reference model. Since the 99% CL flux upper limit is far above the predictions from Earth models due to limited statistics, we do not expect that any realistic Earth models can be used to translate this large flux to heat production. However, if we simply scale the heat production with the relation (9) just to show possible effect that the geoneutrino data would have on geophysical thermal constraints, then the 99% CL flux upper limit is translated to a heat production of 60 TW.

Figure 4 shows the response chart for the KamLAND observation. The observation is in agreement with our Earth model prediction based on the BSE model, although the measurement errors are relatively large. Even if the current KamLAND observation is not as precise as predictions or limits by Earth models, one can see that geoneutrino observation is approaching to the point where we can gain fruitful geophysical information with geoneutrinos.

The current KamLAND geoneutrino observation suffers from large backgrounds originating from reactor anti-neutrinos and $(\alpha,n)$ reactions. The KamLAND group is scheduling further purification of the liquid scintillator needed for solar neutrino detection. The goal is to reduce radioactive contamination at low energy region by a factor of $10^5$ to $10^6$. This purification also benefits geoneutrino detection since it reduces the $(\alpha,n)$ reaction background as well as accidental coincidence background to a negligible level. Reduction of the accidental coincidence background may allow extending the fiducial volume and improving detection efficiencies by relaxing the delayed coincidence event selection criteria.

We estimated prospects of future KamLAND geoneutrino observation after purification by applying the same analysis method as described in [14] to software-generated neutrino event candidates. The extended fiducial volume of 5.5 m radius, together with improved detection efficiency of 90% are used, which are the values currently implemented in the KamLAND reactor neutrino analysis [17]. Figure 5 shows the expected analysis results after 750 day exposure of the detector, which is the same period as used in the current

KamLAND geoneutrino result. KamLAND can determine the geoneutrino flux within 35% accuracy, which is a great improvement from the current accuracy, 54%. The significance of positive geoneutrino signal may reach 99.96%. If we combine it with the current data, the accuracy becomes 28%, which is comparable with the model prediction uncertainty. The 99% flux upper limit may be given at around 30 TW heat generation equivalent, which is below the surface heat-flow measurement results. Separation of $^{238}$U and $^{232}$Th neutrinos seems to be still difficult due to large backgrounds from the surrounding nuclear power reactors, as shown in Figure 5 by the contrast between two contours of with and without the reactor backgrounds.

**Possible sites for future geoneutrino detectors**

From the geophysical and geochemical points of view, one of the most interesting places for geoneutrino observation is Hawaii. Since the Hawaii islands are isolated in the Pacific Ocean, the contribution from the continental crust is negligible, allowing increasing sensitivity to contributions from the lower mantle and the core. Absence of near-by commercial nuclear power reactors is also a great benefit for geoneutrino observation at Hawaii.

Table 3 shows the expected geoneutrino flux at Hawaii calculated with our reference Earth model. The table also shows the flux response coefficients for observation at Hawaii. At Hawaii, the geoneutrino flux from the bulk mantle amounts to 74% of the total flux, and the lower mantle makes the largest contribution of 48%. Due to the large contribution from the mantle, 5 years of geoneutrino observation with a KamLAND class detector ($10^{32}$ target protons, 100% efficiency) at Hawaii can determine the amount of $^{238}$U and $^{232}$Th in the mantle at about 20% accuracy.

Figure 6a shows the cumulative geoneutrino flux as a function of integration radius from a detector, for observation at Hawaii and Kamioka (where KamLAND is located). In contrast to observation at Kamioka, contribution of the crust is very small at Hawaii, and the near-field region does not make significant contribution. Local geology of the Hawaii Islands is fairly simple, since the islands essentially consist of only Oceanic Island Basalt (OIB), whose composition is similar to the oceanic crust. Absence of reactor neutrino backgrounds enables us to observe $^{238}$U and $^{232}$Th geoneutrinos separately, as shown in Figure 5.

For the same reasons, geoneutrino observation at Tahiti is also sensitive to the chemical composition of the lower mantle, and it may be particularly suitable to detect the chemical characteristics of the radioactive element abundances of the Super Plume. This will provide important insights in mantle non-uniformity and a clue to understand the Earth's global dynamics.

Another interesting place for geoneutrino observation is islands on mid-ocean ridges, where new oceanic crust is being formed. Formation of new oceanic crust extracts incompatible elements such as U and Th from the source mantle beneath mid-ocean ridges. Therefore, comparison of the mantle composition between beneath mid-ocean ridges and under the normal oceanic crust gives a hint to understand the long-term mantle evolution and circulation processes. Measuring geoneutrino flux at multiple sites is of further interest. Comparison of the mantle composition beneath mid-ocean ridges of different ages and activities enables us to extract knowledge on chemical evolution of the mantle, crust, and the whole Earth.

**Application of neutrino detectors to study the core energetics**

Recent modeling of the core dynamo energetics suggests that the amount of energy to drive the dynamo is not sufficient without adding radiogenic heat sources such as $^{40}$K, $^{238}$U, and $^{232}$Th in the core (e.g., Buffet [30]). In addition to heat sources such as core solidification heat, some models of the core require 1-10 TW contribution from radiogenic sources.

Although incorporation of K in the core has been argued previously (e.g., Grossmann and Wood [31]), U and Th are also candidates for the energy source in the core, since they are the elements in the early high temperature condensates (e.g. Gessmann [32]). These elements might have been trapped at the base of the mantle during accretion (Anderson [33]), and may form alloys with metallic iron at ultrahigh pressure (e.g., Murell and Burnett [34], Knittle [35]). As shown in Figure 6b, approximately 0.01 to 0.1 ppm of U and Th in the core makes 1 to 10 TW of heat production from the core, and the geoneutrino flux at the Earth surface from this amount of U and Th in the core is about as large as the flux from the mantle. This indicates that if a neutrino detector is positioned in correct places such as Hawaii, it could be one of the most effective probes to explore a possible existence of $^{238}$U and $^{232}$Th in the core, which provides the most essential information on evolution of the core and the magnetic field of the Earth.

**Conclusion**

Neutrinos from $^{238}$U and $^{232}$Th decay within the Earth provide a unique tool to explore the Earth's interior. Observation of these geoneutrinos reveals the chemical constitution of the Earth, providing direct information on the Earth's energy generation mechanism, which is related to understanding planetary formation and evolution.

$7.09 \times 10^{31}$ target-proton-years exposure of KamLAND shows consistent results with our expectation based on the BSE model. Although it is still in the early stage of geoneutrino observation with rather limited statistical power, the radiogenic heat generation upper limit set by the KamLAND observation is comparable to the surface heat-flow measurement. Further accumulation of statistics and planned reduction of radioactive backgrounds will provide more accurate estimation of radiogenic heat generation in the Earth and an independent test to the BSE model. We have demonstrated that geoneutrinos are new potentially very promising probe to explore the Earth's lower mantle and core, which becomes practical in our days.

## Acknowledgement

We are particularly grateful to N. Tolich for useful comments and discussions throughout this work. We also would like to thank A. Kozlov for his help during preparation of this paper. This work is partially supported by the Grant-in-aid of Scientific Research of Priority Area (no. 16075202) of Ministry of Education, Culture, Science, Sport and Technology of Japanese Government to E.O. This work was conducted as a part of the 21st century Center of Excellence programs "Exploring new science by Bridging Particle-Matter Hierarchy" and "Advanced Science and Technology Center of the Dynamic Earth" of Tohoku University.

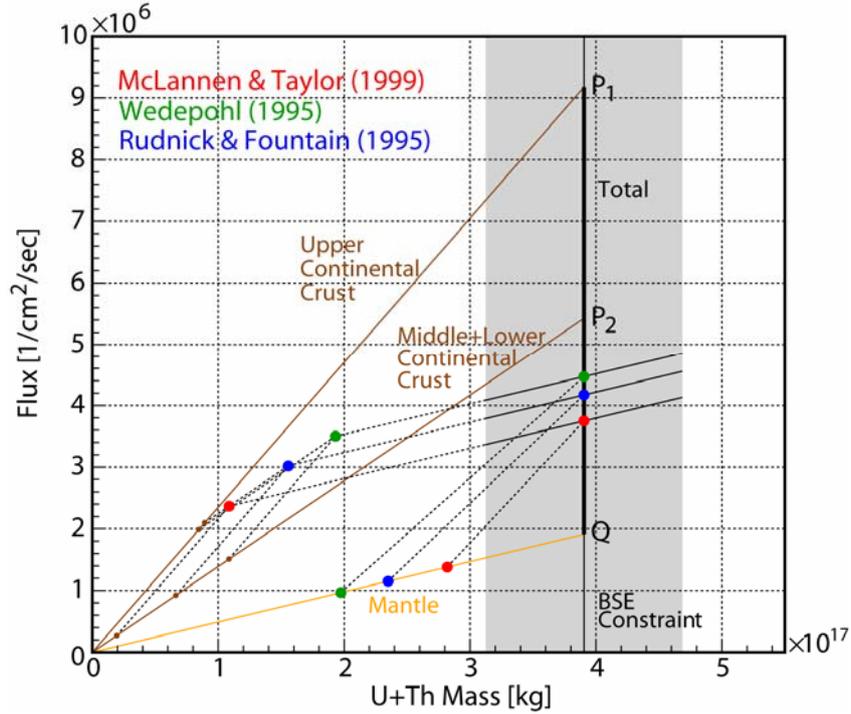

**Figure 1**: Neutrino flux dependence on crustal models. The dark brown lines (OP$_1$ and OP$_2$) are the response lines of the upper continental crust and middle + lower continental crust. Since some of crustal models do not separate the middle and lower crust, and since they have similar response coefficients, these are considered together. The yellow line (OQ) is the response line of the mantle. The thick black vertical line represents the central value of the BSE constraint, and the shaded area shows the range of total U and Th mass allowed assuming a 20% error in the BSE constraint. The point P$_1$ represents an extreme case that all U and Th are dispersed in the upper continental crust, and the point Q represents another extreme case that all U and Th are dispersed in the mantle. Between the response lines of the crust, three representative crustal composition estimations with relation to the total flux, i.e., McLannen & Taylor (1999) [36], Wedepohl (1995) [37] and Rudnick & Fountain (1995) [19], and mantle as a residual from the BSE composition are shown. Each of the three thin solid black lines in the shaded area shows the range of the total flux variation due to uncertainty in BSE estimation under a given crustal model. In this example, we fix the crustal composition and the variation of the bulk composition is translated to variation of the mantle composition.

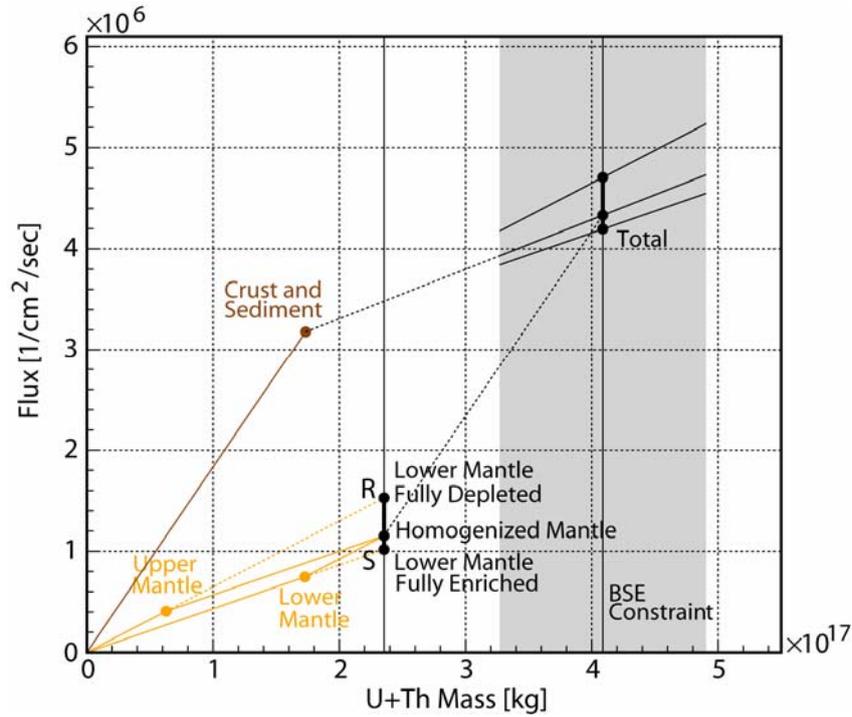

**Figure 2**: Neutrino flux dependence on mantle models. The shaded area and the vertical line in the shaded area show the BSE constraint and its error, respectively, and the left vertical line shows the amount of U and Th in the mantle calculated from the BSE constraint and the adopted crustal model. The left vertical thick line (RS) shows the possible variation of the flux from the mantle among different distributions of fixed amount of U and Th between the upper and lower mantle. The points R and S represent two extreme cases, namely the fully depleted lower mantle and the fully enriched lower mantle, respectively. Our reference model adopts homogenized mantle as a starting point, which is represented by the point between R and S. Each of the three diagonal lines in the shaded area shows the ranges of flux variation due to BSE error, for each mantle model. In this example, the variation of the bulk composition is translated to the mantle composition without changing the distribution ratio between the upper and lower mantle.

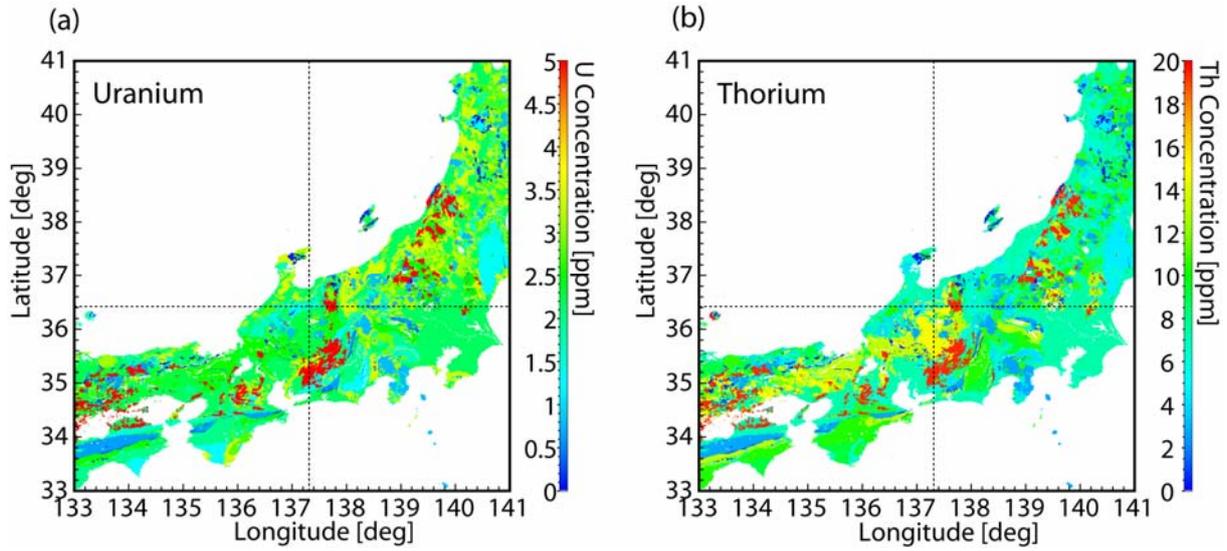

**Figure 3**: A local surface geology model. U and Th distribution maps are constructed by combining a geological map [27] and a survey of the Japanese Islands geochemistry based on the geological map [23]. The geological map is made mainly based on steep V-shaped valley outcrops, thus it represents the base rocks. Broad similarities to a geochemical map [38], which is made based on surface river sediment, supports appropriateness of this method, although river sediment shows overall tendency of lower U and Th concentrations than those of base rock samples.

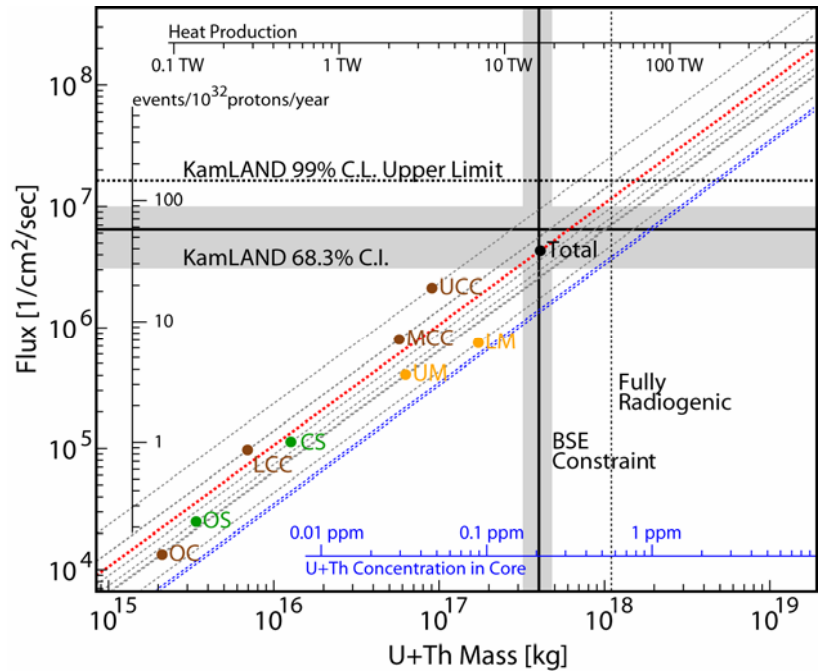

**Figure 4**: KamLAND observation and the reference Earth model. The horizontal line is the KamLAND best-fit flux and the horizontal shaded band shows the interval of 68.3% C.L. The horizontal dotted line is the 99% C.L. upper limit. The points represent the expected neutrino flux at the KamLAND site in the reference Earth model; Upper Continental Crust (UCC), Middle Continental Crust (MCC), Lower Continental Crust (LCC), Oceanic Crust (OC), Continental Sediment (CS), Oceanic Sediment (OS), Upper Mantle (UM) and Lower Mantle (LM). This assumes a Th/U ratio of 3.9. The blue diagonal lines are response lines of the inner core (lower line) and the outer core (upper line). In the reference model, U and Th amount in the core is set to be zero. The blue auxiliary axis at the bottom indicates the concentration of the U+Th in the core assuming a Th/U ratio of 3.9, for a given U+Th mass in the core. The black auxiliary axis on the top shows heat generation from U and Th for a given U+Th mass.

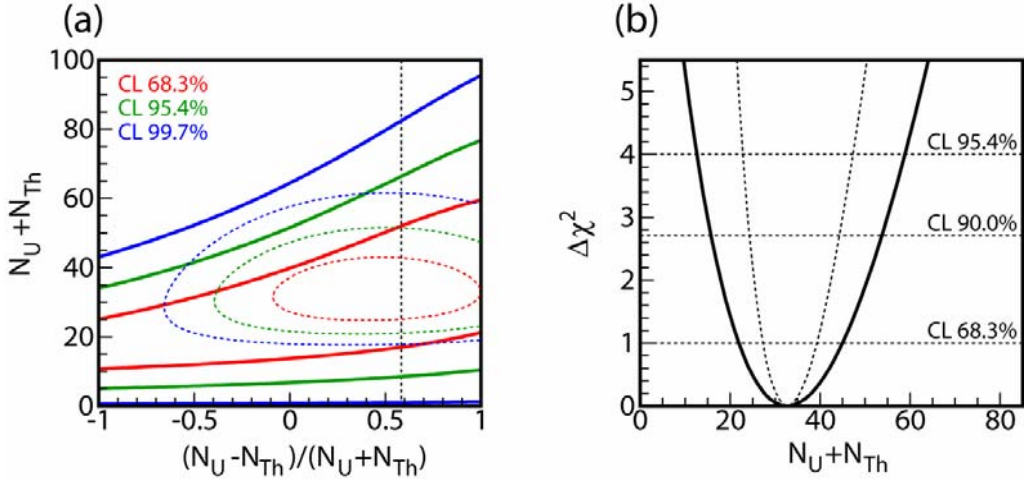

**Figure 5**: Prospects of future KamLAND geoneutrino observation after purification. The same analysis method as described in [14] is applied, except for reduced radioactive background estimation and consequent selection criteria relaxation. The solid lines in the panel (a) show the 68.3% CL (red), 95.4% CL (green), and 99.7% CL (blue) contour lines for the total U and Th neutrinos, $N_U + N_{Th}$, versus the normalized difference, $\frac{N_U - N_{Th}}{N_U + N_{Th}}$. The vertical dotted line represents a constraint given by a cosmochemical analysis, Th/U = 3.9. The solid line in the panel (b) shows $\Delta\chi^2$ as a function of total number of U and Th geoneutrino events if we fix the U/Th ratio to the cosmochemical constraints. The same analysis of prospects except for absence of reactor neutrino background is also shown with the dotted lines in both plots. Absence of reactor backgrounds improves not only the absolute rate estimation but also the U and Th separation capability.

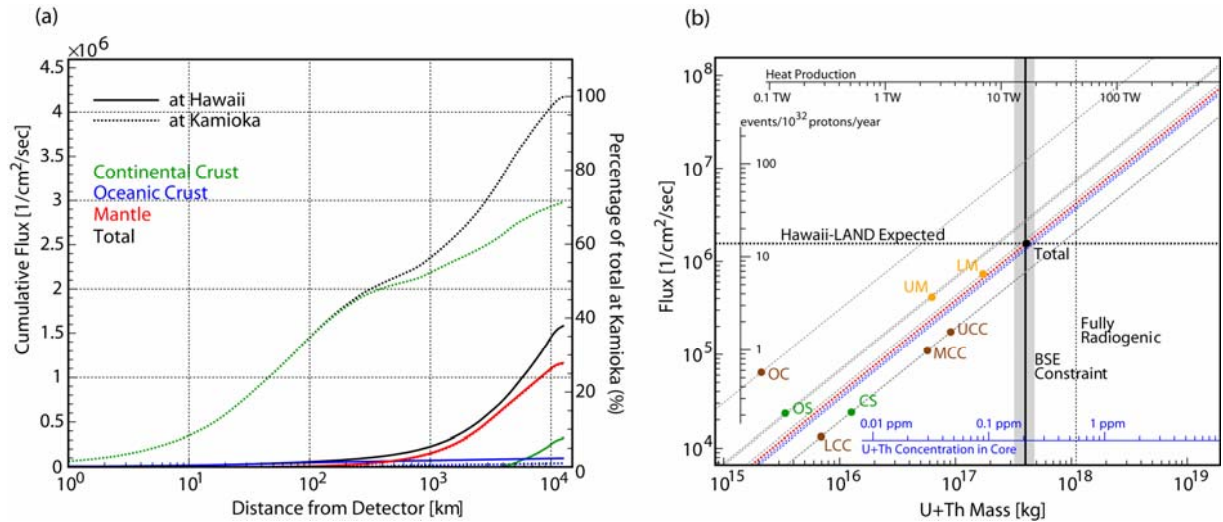

**Figure 6**: Cumulative geoneutrino flux within a given distance from a detector, for observation at Hawaii and Kamioka, and the flux response chart for observation at Hawaii. As shown in panel (a), at Hawaii, in contrast to Kamioka, contributions from the mantle and core are very large, whereas those from near-field region are very small. Panel (b) shows that possible U and Th contained in the core (1-10TW) make detectable contribution to the total flux. Also as shown in panel (b), small differences in the response coefficients among UCC, MCC, LCC and CS indicate that the neutrino flux at Hawaii is independent from the U and Th distribution within the continental crust. Although the continental crust contributes about 25% of the total flux at Hawaii, this fact reduces one of the largest uncertainties significantly, and realizes robustness for study of deep portion of the Earth.

**Table 1**: The reference Earth model.

| Reservoir | | Concentration [ppm] | | Mass ×10$^{16}$ [kg] | | Heat Production [TW] | |
|---|---|---|---|---|---|---|---|
| | | U | Th | $^{238}$U | $^{232}$Th | $^{238}$U | $^{232}$Th |
| Sediment | Continental | 2.8 | 10.7 | 0.26 | 0.99 | 0.26 | 0.26 |
| | Oceanic | 1.7 | 6.9 | 0.07 | 0.28 | 0.07 | 0.07 |
| Continental Crust | Upper | 2.8 | 10.7 | 1.85 | 7.08 | 1.75 | 1.86 |
| | Middle | 1.6 | 6.1 | 1.17 | 4.47 | 1.11 | 1.17 |
| | Lower | 0.2 | 1.2 | 0.14 | 0.85 | 0.13 | 0.22 |
| Oceanic Crust | | 0.10 | 0.22 | 0.04 | 0.09 | 0.04 | 0.02 |
| Mantle | Upper | 0.012 | 0.048 | 1.28 | 5.13 | 1.21 | 1.35 |
| | Lower | 0.012 | 0.048 | 3.52 | 14.1 | 3.32 | 3.71 |
| Core | Outer | 0 | 0 | 0 | 0 | 0 | 0 |
| | Inner | 0 | 0 | 0 | 0 | 0 | 0 |
| Bulk Silicate Earth (BSE) | | 0.0203 | 0.0795 | 8.18 | 32.1 | 7.73 | 8.44 |

**Table 2**: Geoneutrino fluxes for observation at the KamLAND site, calculated with the reference Earth model. The flux response coefficients are independent from geochemical modeling in the reference Earth model (i.e., U and Th mass in each reservoir).

| Reservoir | | Neutrino Flux $\times 10^5$ [1/cm$^2$/sec] | | Response Coefficient $\times 10^{-12}$ [1/cm$^2$/sec / kg] | |
|---|---|---|---|---|---|
| | | $^{238}$U | $^{232}$Th | $^{238}$U | $^{232}$Th |
| Sediment | Continental | 0.61 | 0.51 | 23.6 | 5.13 |
| | Oceanic | 0.14 | 0.12 | 19.5 | 4.24 |
| Continental Crust | Upper | 11.5 | 9.57 | 62.2 | 13.5 |
| | Middle | 4.31 | 3.57 | 36.8 | 8.00 |
| | Lower | 0.53 | 0.69 | 37.1 | 8.06 |
| Oceanic Crust | | 0.09 | 0.04 | 16.8 | 3.65 |
| Mantle | Upper | 2.20 | 1.91 | 17.2 | 3.74 |
| | Lower | 4.03 | 3.51 | 11.4 | 2.49 |
| Core | Outer | 0 | 0 | 9.27 | 2.02 |
| | Inner | 0 | 0 | 8.73 | 1.90 |
| Bulk Silicate Earth | | 23.4 | 19.9 | 29.0 | 6.17 |

**Table 3**: Geoneutrino flux and flux response coefficients for observation at Hawaii, calculated with the reference Earth model.

| Reservoir | | Neutrino Flux $\times 10^5$ [1/cm²/sec] | | Response Coefficient $\times 10^{-12}$ [1/cm²/sec / kg] | |
|---|---|---|---|---|---|
| | | $^{238}$U | $^{232}$Th | $^{238}$U | $^{232}$Th |
| Sediment | Continental | 0.13 | 0.11 | 5.03 | 1.09 |
| | Oceanic | 0.13 | 0.11 | 18.4 | 3.99 |
| Continental Crust | Upper | 0.94 | 0.78 | 5.08 | 1.11 |
| | Middle | 0.60 | 0.49 | 5.09 | 1.11 |
| | Lower | 0.07 | 0.09 | 5.09 | 1.11 |
| Oceanic Crust | | 0.43 | 0.21 | 80.5 | 17.5 |
| Mantle | Upper | 2.23 | 1.94 | 17.5 | 3.80 |
| | Lower | 4.03 | 3.51 | 11.4 | 2.49 |
| Core | Outer | 0 | 0 | 9.27 | 2.02 |
| | Inner | 0 | 0 | 8.73 | 1.90 |
| Bulk Silicate Earth | | 8.57 | 7.25 | 10.3 | 2.20 |